\documentclass[10pt]{article}

\usepackage[margin=0.8in]{geometry}
\usepackage{setspace}
\usepackage{amsmath,amssymb,amsthm}
\usepackage{graphicx}
\usepackage{multicol}
\usepackage{flafter}
\usepackage{authblk}
\usepackage{color,hyperref,xcolor}
\hypersetup{colorlinks=true,urlcolor=blue, citecolor=purple}
\usepackage{cleveref}

\usepackage[round]{natbib}
\numberwithin{equation}{section}

\def\R{$\mathcal{R}_0$ }

\begin{document}

\title{A time series method to analyze incidence pattern and estimate reproduction number of COVID-19}

\author{
 Soudeep Deb
 \thanks{Email: soudeep@iimb.ac.in; Corresponding author}}
 \affil{Indian Institute of Management, Bangalore \\ Bannerghatta Main Rd, Bilekahalli \\ Bengaluru, Karnataka 560076, India.}

\author{
 Manidipa Majumdar
 \thanks{Email: manidipa.majumdar@yahoo.in}}
 \affil{Sri Jayadeva Institute of Cardiovascular Science and Research \\ Bannerghatta Main Rd, Jayanagar 9th Block \\ Bengaluru, Karnataka 560069, India.}

\date{}

\maketitle

\begin{abstract}
\noindent The ongoing pandemic of Coronavirus disease (COVID-19) emerged in Wuhan, China in the end of 2019. It has already affected more than 300,000 people, with the number of deaths nearing 13000 across the world. As it has been posing a huge threat to global public health, it is of utmost importance to identify the rate at which the disease is spreading. In this study, we propose a time series model to analyze the trend pattern of the incidence of COVID-19 outbreak. We also incorporate information on total or partial lockdown, wherever available, into the model. The model is concise in structure, and using appropriate diagnostic measures, we showed that a time-dependent quadratic trend successfully captures the incidence pattern of the disease. We also estimate the basic reproduction number across different countries, and find that it is consistent except for the United States of America. The above statistical analysis is able to shed light on understanding the trends of the outbreak, and gives insight on what epidemiological stage a region is in. This has the potential to help in prompting policies to address COVID-19 pandemic in different countries.
\end{abstract}


\newpage

\section{Introduction}
\label{sec:introduction} 

\subsection{Context}

In December 2019, following reports of a cluster of cases resembling viral pneumonia, associated with a point source of spread from a fish market in Wuhan, Hubei, China, the search for identifying the causative agent and retarding the spread of an ensuing epidemic began. COVID-19, as named by World Health Organization (WHO) on 11th February 2020, represents the newest zoonotic Coronavirus disease that crossed species to affect humans and spread in an unprecedented manner. As of 21st March, 2020, more than 300,000 cases have been confirmed across 166 countries in six continents, causing more than 13,000 deaths. The outbreak was declared a pandemic (involving more than 3 geographical regions), a Public Health Emergency of International Concern on 30 January 2020, by WHO. 

This outbreak is the third major human Coronavirus epidemic in the last 2 decades, after the SARS (Severe Acute Respiratory Syndrome) in 2002-2003 which caused more than 8000 cases with around 9-10\% fatality rate and MERS (Middle East Respiratory Syndrome) in 2015 which recorded more than 2500 cases, with a fatality rate of 34\%. Unlike those two epidemics, this time the healthcare response was quicker and more robust, though high infectivity and international travel probably led to a greater transmission. Within a month of the first reported case, on 7th Januray, 2020, novel Coronavirus (2019-nCoV) was isolated by deep gene sequencing from lower respiratory tract samples of patients and its sequence was published on Jan 12, 2020. This enabled development of point of care RT-PCR tests and IgG and IgM ELISA tests specific for 2019-nCoV very soon after first reporting of cases. These tests are based on full genome sequence data on the Global Initiative on Sharing All Influenza Data (GISAID).

\subsection{Description of the disease}

Case definitions were brought forth by WHO and different national health authorities and a confirmed case was defined as a case with respiratory specimens (commonly broncho-alveolar lavage) which tested positive for the 2019-nCoV in at least one of the following three methods: (a) isolation of 2019-nCoV, (b) at least two positive results by real-time, reverse-transcription–polymerasechain-reaction (RT-PCR) assay for 2019-nCoV, or (c) a genetic sequence that matches 2019-nCoV.

2019-nCoV closely resembles the SARS-CoV (about 80\% sequence identity, and uses the same cell entry receptor -ACE-II), MERS-CoV and a bat coronavirus (with which it has 96\% sequence identity) which are enveloped non-segmented positive sense RNA viruses from the family Coronaviridae broadly distributed in mammals, and is likely to behave clinically and epidemiologically akin to SARS. Initial transmission of the disease involved seafood to human (zoonotic) transmission, which rapidly evolved into human to human transmission in secondary and tertiary cases. Human to human transmission of 2019-nCoV  occurs through large droplets and contact and less so by means of fomites and aerosols. It has also been seen to survive for days on surfaces like cardboard, steel and plastic. \cite{li2020early} described the early transmission dynamics of the disease among initially affected Chinese population. Their preliminary estimate of the incubation period distribution indicates a 14-day medical observation period or quarantine for exposed persons. The rapidity with which the virus has been noted to mutate, as evidenced by ability of human to human transmission, it is unlikely that a specific effective vaccine can be developed soon. Thus epidemiological interventions like lock-down or mass screening and quarantine might be more effective to contain the spread.
 
\subsection{Literature Survey}

There have already been a lot of studies on this recent epidemic. They cover a wide range of aspects. \cite{huang2020clinical} described the spectrum of clinical presentation of 41 Chinese patients from Wuhan. The commonest symptoms were fever (98\%), cough (76\%) and myalgia (44\%), while sputum production, headache and diarrhoea were rare presentations. Acute respiratory distress syndrome (ARDS), anemia, secondary infection, and acute cardiac injury were common complications noted more in patients needing intensive care. \cite{wang2020clinical} described clinical characteristics of 138 cases of NCIP (novel Coronavirus infected pneumonia) admitted in Zhongnan Hospital of Wuhan. They reiterated that fever, fatigue and dry cough were the commonest symptoms. Majority of the patients were in their fifth decade. 26\% patients needed intensive care, with ARDS, shock and arrhythmia being the major complications. 
 
The epidemic has been second most severe in Italy so far. \cite{albarello20202019} has discussed two of the cases from Italy which helped understanding some diagnostic aspects of the disease. \cite{holshue2020first} reported the first case of 2019-nCoV in the United States.
 
\cite{li2020therapeutic} explored therapeutic options with repurposing of older approved (favipiravir and ribavirin) or experimental antivirals (remdesivir and galidesivir) in view of highly conserved genetic sequence of the cat6alytic sites of the four 2019-nCoV enzymes which have high sequence similarity with those of SARS and MERS Coronaviruses. Baloxavir marboxil (influenza inhibitor of the cap-dependent endonuclease), interferons, ribavirin, Griffithsin (a red-alga-derived viral glycoprotein binding lectin), Lopinavir and Ritonavir (approved anti HIV drug), Nitazoxanide, Chloroquine are few of the drugs under scrutiny for use in COVID-19. \cite{zhu2020systematic} very recently published a systematic review of all ongoing clinical drug trials relating to this. Among many drugs studied so far, hydroxychloroquine, though pending FDA approval, appears promising and its use is supported by multiple studies like \cite{gao2020breakthrough} and \cite{cortegiani2020systematic}. Though development of targeted vaccine in this mutating virus may be effective in halting future epidemics, their use in the current emergent scenario seems to be limited, as it may take months or years to develop an effective, widely tested vaccine.

The epidemic spreading wildly, there have been a surge of statistical studies to predict the number of cases affected by the virus. \cite{anastassopoulou2020data}, \cite{gamero2020forecast} and \cite{wu2020nowcasting} are a few notable examples in this regard. Elsewhere, \cite{lin2020conceptual} researched on the effect of individual reaction and governmental action (holiday extension, city lockdown, hospitalisation and quarantine) on the spread of the infection. \cite{chinazzi2020effect} studied the effect of travel restrictions on the same. 

\subsection{Our contribution}

We, in this study, work with an interesting aspect of the COVID-19 outbreak. We develop a novel mathematical model to study the trend pattern of the incidence, defined by the number of new cases identified everyday, for different provinces in China and for six different countries. The implication of our work is three-fold. First, it provides insight on the rate of growth of the number of affected people. In particular, we discovered that a quadratic trend was able to explain the data best. Second, we estimate the time when there is a significant change in the growth pattern of the disease. For China, both on country-level and on province-level, it was found that the growth ceased towards the end of February. In contrast, the other countries are seeing a surge in the growth rate from around the same time. The results from this analysis can help one identify possible factors which directly or indirectly dictate the change in the growth pattern of COVID-19 incidence, for better (enabling similar strategy initiation in different population or early in the same population), or for the worse (which might indicate challenges that are to be prioritised). Finally, we estimate the basic reproduction number for different countries using the fitted values of our model. It was found that the estimates are consistent with the range provided by WHO.

\section{Preliminaries}
\label{sec:preliminaries}

\textbf{Data source:}

All data related to COVID-19 were collected from \cite{datarepository}, the GitHub repository provided and maintained continuously by the Johns Hopkins University Center for Systems Science and Engineering (JHU CSSE). We also use population data, both on province-level for China and on country-level for the world. These were collected respectively from \cite{data-chinapopulation} and \cite{data-globalpopulation}.

\textbf{Variables:}

The data consist of three main variables - number of confirmed cases, number of deaths and number of people recovered in different regions. Throughout the paper, they will be denoted by $y_{k,t}$, $d_{k,t}$ and $r_{k,t}$, respectively. In a similar fashion, population is denoted by $P_{k,t}$. Here, $k$ is used for a region and $t$ is a time-point. The data is provided on daily level, starting from 22nd January, 2020 to 21st March, 2020. 

In our analysis, we also study the effect of lockdowns on the growth rate of the number of affected cases. The Chinese province of Hubei went into complete lockdown on 23rd January, 2020. A few areas in other provinces also went to partial lockdown, and observed something called "closed management". This means that those areas would only keep one entrance and exit point open, and each household is allowed limited numbers of entrances and exits per day. In addition, night-time access was prohibited in some areas as well. Our method is able to quantify the effect of lockdowns (full or partial) in respective provinces, once we account for the general time trend. For that, we create a binary variable $L_{k,t}$ to denote pre (corresponds to $L_{k,t}=0$) and post (corresponds to $L_{k,t}=1$) lockdown periods.

Any other variable used in the analysis will be defined later, as needed.

\textbf{Definitions:}

Throughout this paper, we are primarily interested in the prevalence and incidence of the disease. Prevalence, in epidemiology, is defined as the proportion of people in a population affected by the disease. In this study, we calculate it as the number of cases per 1,000,000 people. Thus, if $y_{k,t}$ and $P_{k,t}$ denote respectively the number of affected cases and population (in millions) of region $k$ at time $t$, then prevalence ($\pi_{k,t})$ is computed as 
\begin{equation}
\label{eqn:prevalence}
\pi_{k,t} = \frac{y_{k,t}}{P_{k,t}}.
\end{equation}

On the other hand, incidence is defined as the proportion of new cases per 1,000,000 people. With the same variables as above, incidence ($\theta_{k,t})$ is computed as 
\begin{equation}
\label{eqn:incidence}
\theta_{k,t} = \frac{y_{k,t}-y_{k,t-1}}{P_{k,t}}.
\end{equation}

Another contribution of this paper is to provide country-wise estimates for the basic reproduction number, hereafter denoted by $\mathcal{R}_0$. In epidemiology, \R of an infection is the expected number of cases directly caused by one affected case when all individuals in the population are equally susceptible to infection.

In order to estimate $\mathcal{R}_0$, we use the serial interval ($\mathcal{I}$) as well. $\mathcal{I}$, in epidemiology of infectious diseases, refers to the time between successive cases in a chain of transmission. One can estimate it by the time interval between infection and subsequent transmission. Thus, if the timing of infection transmission in deference to a person's clinical onset is $T_1$, and the incubation period of a subsequent case is $T_2$, then 
\begin{equation}
\label{eqn:serial-interval}
    \mathcal{I} = T_1+T_2
\end{equation}

\section{Methods}
\label{sec:methods}

\subsection{Model}

For most parts of the statistical analysis in this paper, we use auto-regressive integrated moving average (ARIMA) methods with time-dependent parameters. The goal of this study is to identify if the coefficient estimate for the time trend changes after a certain point, and if so, we examine it further to find out the nature of change and potential causes. Our model is defined below using the notations from last section, but we drop the subscript $k$ for simplicity.

Formally, let $\theta_t$ denote the incidence rate (see \cref{eqn:incidence}) of COVID-19 at time $t \in \{1,2,\hdots,T\}$. $P_t$ denote the overall population (in millions). $L_t$ is the dummy variable to signify lockdown. As mentioned earlier, every time point $t$ denotes a single day. We shall use $\tau$ to denote the day when the first confirmed case is observed, i.e. $y_t = 0$ for $t<\tau$ and $y_t > 0$ for $t\ge \tau$. In the data, it is observed that the number of affected cases (denoted by $y_t$) grows with time, more than linearly at first and then stabilizing after a certain time. Keeping that in mind, we consider the following structure for the logarithm of $\theta_t$.
\begin{equation}
\label{eqn:model-structure}
    \log \theta_t := f(t,\tau) + \gamma L_t + u_t.
\end{equation}

Here, $f(t,\tau)$ is the trend function, $\gamma$ captures the effect of lockdown, and $u_t$ is the error process. We want to point out that one can include more additive terms in the mean structure to analyze effects of other covariates (for example, policy change, healthcare system etc), if relevant information is available. In \cref{eqn:model-structure}, we assume an ARMA structure for the error process while $f$ is assumed to be polynomial in nature. The maximum degree for the polynomial function was considered to be 3 at first. But, experimentation with the data suggested that a quadratic trend in $(t-\tau)$ is best suited for the incidence model, across different regions. This quadratic function is defined in a time-dependent way. In particular, the coefficients for the linear and quadratic terms in $f$ are considered to be different for $(t-\tau)<\eta$ and $(t-\tau)\ge \eta$. The value of $\eta$ is crucial in this study. It is estimated from the data and it tells us when the trend of the growth changes its pattern. The corresponding coefficient estimates then provide an idea about how the growth is changing, for better or for worse. Combining everything, following is the equation of the model we use:
\begin{equation}
\label{eqn:model}
    \log \theta_t = \beta_{0} + \beta_{1,t}(t-\tau) + \beta_{2,t}(t-\tau)^2 + \gamma L_t + \sum_{j=1}^p\phi_j\log \theta_{t-j} + \epsilon_t + \sum_{j=1}^q\gamma_j\epsilon_{t-j}.
\end{equation}
Here, $p$ and $q$ denote the auto-regressive (AR) and moving-average (MA) orders of the error process $u_t$. $\epsilon_t$ denotes a standard normal white noise process and for $m=1,2$,
\begin{equation}
\beta_{m,t} = \left\{ \begin{array}{cc} \beta_{m,1} & \text{for $t<\eta$} \\ \beta_{m,2} & \text{for $t\ge\eta$} \end{array}\right..
\end{equation}

Throughout the paper, we estimate $\eta, p, q$ from the data using Akaike information criteria (AIC).

\subsection{Diagnostics}

After fitting the model, we run a few residual diagnostics. First, a Box-Ljung test is performed on the residuals to test if any of the autocorrelation is significantly different from zero. These tests were done at 5\% level of significance. Fitted vs residual plot, QQ plot of the residuals, partial autocorrelation function (PACF) plot of the residuals are also explored to confirm that the model is appropriate for the data.

Next, predictive ability of the model is evaluated. For every site, we use all but last three days' data to train the model, and use the estimated coefficients to predict the log-incidence of those three days. The prediction, hereafter denoted as $\widehat{\log\theta_t}$, is then transformed to incidence and subsequently the number of new cases for that day is computed. Root mean squared error (RMSE), as defined by \cref{eqn:rmse}, is finally calculated to understand how well the model can forecast the future. 
\begin{equation}
\label{eqn:rmse}
    \textrm{RMSE} = \sum \{(P_{t}\exp(\widehat{\log\theta_t}) - (y_{t}-y_{t-1})\}^2.
\end{equation}

\subsection{Estimating reproduction number}
\label{subsec:estimate-r0}

For every country, we use the number of new cases per day from our model and use that to estimate \R using maximum likelihood method. Note that we need to use the distribution of the serial interval ($\mathcal{I})$ in this process. Unfortunately, that is still not known for 2019-nCoV. So, similar to \cite{zhao2020preliminary}, we use different distributions of $\mathcal{I}$ to estimate the reproduction number. There are three candidate distributions we have used in this study: (a) same as SARS (gamma distribution, with mean 8.4 and standard deviation 3.8), (b) same as MERS (gamma distribution, with mean 7.6 and standard deviation 3.4), and (c) average of SARS and MERS (gamma distribution, with mean 8.0 and standard deviation 3.6). 

\subsection{Implementation}

All of the above methods are implemented using RStudio version 1.2.5033, along with R version 3.6.2. All ARIMA related computations are done using the {\sf{forecast}} package. The estimation of \R was performed with the help of {\sf{R0}} package.

\section{Results}
\label{sec:results}

We apply the above methods in two parts. First, we focus on the province-level daily data from China and observe how the trend of the incidence changes in different parts of the country. These results are described in \Cref{subsec:china}. Next, we consider the country-level daily data from China, India, Iran, Italy, South Korea and the United States of America; and discuss the findings in \Cref{subsec:global}.

\subsection{Province-level analysis for China}
\label{subsec:china}

There are 31 Chinese provinces affected by COVID-19. Maximumn number of affected cases, as expected, is in Hubei. Till date, there have been 67800 reported cases in Hubei, among which 3139 (4.63\%) died. Next highest number of reported cases is in Guangdong (1400), although the number of deaths is only 8 (0.57\%) there. Tibet and Qinghai have seen the lowest number of confirmed case, with 1 and 18 respectively. Their population are also the lowest of the country. On that note, prevalence, defined by \cref{eqn:prevalence}, is a better measure to understand the extent of outbreak. Hubei has observed a prevalence of staggering 1145.85 cases per million people. Next highest on the list is Beijing (23.40), followed by Zhejiang (21.54) and Jianxi (20.12). So far as the fatality is concerned, other than Hubei, Xinjiang (3.95\%) and Hainan (3.57\%) are the other two provinces to have a more than 3\% rate. Recovery rate is more than 75\% for all provinces. 

\begin{figure}[!hbt]
\begin{center}
\includegraphics[width=\textwidth,keepaspectratio]{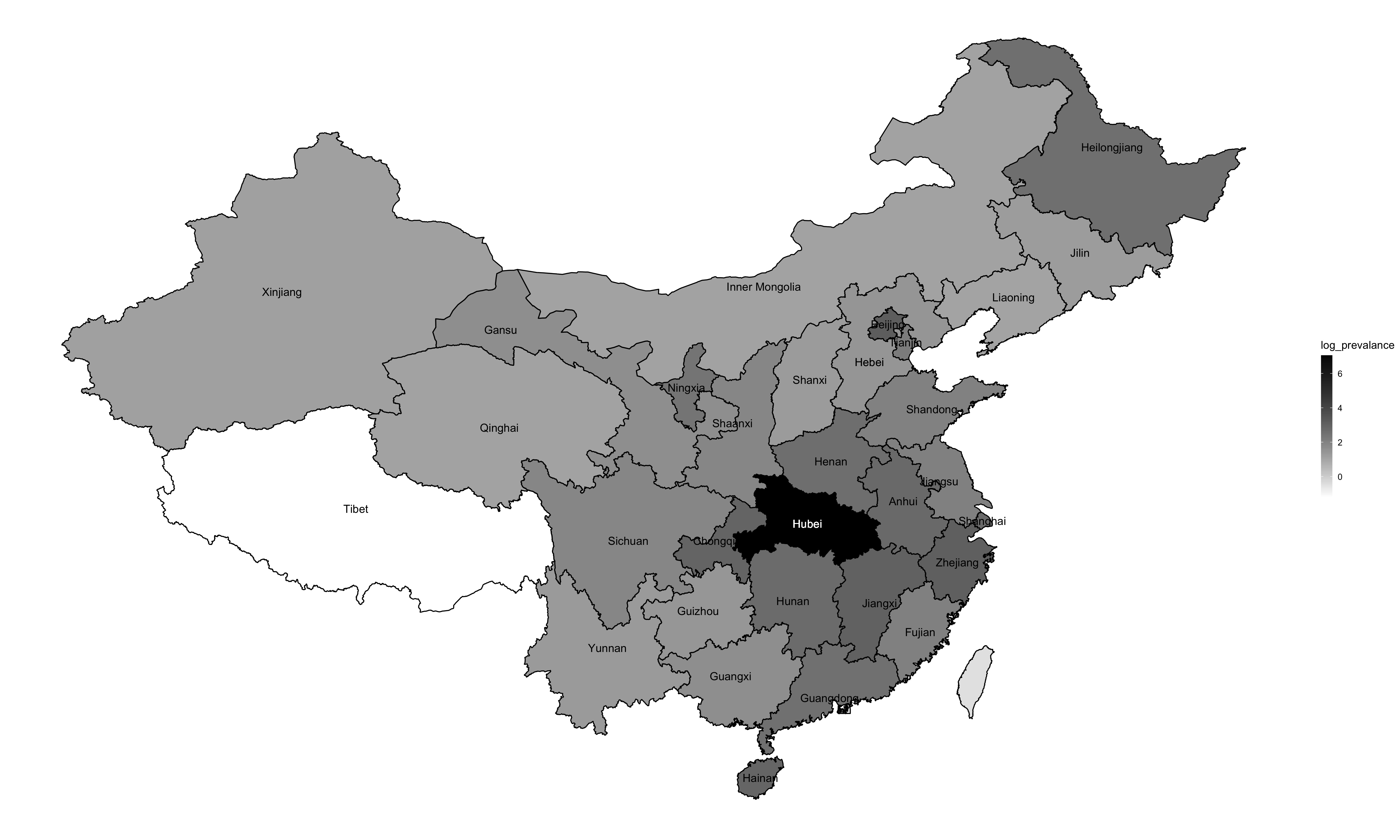}
\end{center}
\caption{Heatmap of the log-prevalence for different provinces in China.}
\label{fig:heatmap-china}
\end{figure}

Overall, there have been 81305 confirmed cases in China, among which 4.01\% have died and 88.38\% have recovered. A heatmap of the whole country based on the prevalence (changed to logarithmic scale) is shown in \Cref{fig:heatmap-china}. Hubei is a clear outlier. It is also evident from the map that the provinces closer to Hubei were usually affected more than the others. 

Model (\ref{eqn:model}) is now applied to the log-incidence data of every province separately. Note that the length of every time series is 60, as we are dealing with daily data from 22nd January, 2020 to 21st March, 2020. As mentioned in \Cref{sec:methods}, AIC is used to identify the date when the trend pattern of the log-incidence changes. Most suitable AR and MA orders were also chosen accordingly. In \Cref{tab:results-china}, we present the most appropriate ARMA order, effect of lockdowns (partial or full) wherever appropriate, p-value from the Box-Ljung test on residuals and RMSE (on the number of new-cases) for three-day-prediction for the best models corresponding to different Chinese provinces.

\begin{table}[!htb]
\centering
\caption{Results and diagnostics for daily data from all Chinese provinces}
\label{tab:results-china}
\begin{tabular}{lcccc}
  \hline
  Province & ARMA order & lockdown-effect & p-value (Box test) & RMSE (prediction) \\ 
  \hline
  Anhui & (2,1,1) &  & 0.72 & 0.53 \\ 
  Beijing & (0,1,1) & 9.47 (1.3)* & 0.76 & 2.75 \\ 
  Chongqing & (1,1,0) & 11.35 (1.3)* & 0.97 & 0.59 \\ 
  Fujian & (1,1,2) & -0.1 (1.7) & 0.82 & 2.59 \\ 
  Gansu & (0,1,0) & -3.17 (2) & 0.78 & 0.47 \\ 
  Guangdong & (0,1,1) & 35.1 (4.2)* & 0.68 & 7.35 \\ 
  Guangxi & (0,1,1) & -2.31 (0.7)* & 0.59 & 0.52 \\ 
  Guizhou & (0,1,1) & 1.32 (1) & 0.88 & 0.57 \\ 
  Hainan & (3,1,0) & 1.02 (0.6) & 0.40 & 0.11 \\ 
  Hebei & (1,1,0) & 0.2 (1.4) & 0.77 & 0.62 \\ 
  Heilongjiang & (0,1,0) & -1.22 (1.1) & 0.87 & 0.54 \\ 
  Henan & (1,1,0) & 0.21 (1.1) & 0.22 & 0.80 \\ 
  Hubei & (0,1,0) & 6.67 (1.3)* & 0.64 & 1.87 \\ 
  Hunan & (2,1,0) &  & 0.82 & 0.56 \\ 
  Inner Mongolia & (0,1,1) & 4.43 (1.2)* & 0.83 & 0.30 \\ 
  Jiangsu & (1,1,2) & -0.11 (0.3) & 0.98 & 0.73 \\ 
  Jiangxi & (3,1,0) & -0.38 (0.2) & 0.87 & 0.13 \\ 
  Jilin & (2,1,2) & 0.32 (1.2) & 0.74 & 0.23 \\ 
  Liaoning & (0,1,0) & -0.88 (1.4) & 0.28 & 0.49 \\ 
  Ningxia & (0,1,0) & -5.49 (2.5)* & 0.19 & 0.06 \\ 
  Qinghai & (6,1,0) &  & 0.90 & 0.06 \\ 
  Shaanxi & (2,1,0) & 0 (0.7) & 0.88 & 0.38 \\ 
  Shandong & (0,1,0) & 24.62 (2.6)* & 0.71 & 0.82 \\ 
  Shanghai & (0,1,0) & 1.51 (2.6) & 0.40 & 5.28 \\ 
  Shanxi & (0,1,0) &  & 0.40 & 0.29 \\ 
  Sichuan & (1,1,0) &  & 0.89 & 0.53 \\ 
  Tianjin & (0,1,0) & -0.89 (1.3) & 0.72 & 0.51 \\ 
  Tibet & (0,1,0) &  & 0.71 & 0.00 \\ 
  Xinjiang & (1,1,0) &  & 0.36 & 0.27 \\ 
  Yunnan & (0,1,0) & -3.36 (1.3)* & 0.94 & 0.62 \\ 
  Zhejiang & (0,1,0) & -1.66 (1.8) & 0.63 & 0.42 \\ 
   \hline
\end{tabular}
\end{table}

The results show that differencing was necessary for all provinces, thereby suggesting that all of the time series were nonstationary. For Qinghai, which has observed one of the lowest prevalence rates, the most appropriate ARMA order was found to be $(6,1,0)$. This describes a significant dependence with up to lag 6, that is the previous 6 days. All other provinces showed dependence with at most prior 3 days. The last two columns of \Cref{tab:results-china} report diagnostics results. The p-value of the Box-Ljung test on the residuals was more than 0.05 for all provinces. It establishes that the residuals are not autocorrelated, thereby showing the adequacy of the model to capture the time dependence pattern appropriately. On the other hand, RMSE of the predictions of new cases in a three-day-ahead time window was at most 7.35. Hence, the predictive ability of the model is good, so far as the Chinese provinces are concerned.

In the above table, we also present the effect of lockdown (partial or full) in provinces who imposed such rules. It was found to be significant only in a few provinces, and some of them still showed positive effect. This might be attributed to the theory that lockdown only ensured no exposure, but a large proportion of population were already affected and got tested thereafter.

The date on which the trend pattern of the log-incidence series changes significantly is one of the most interesting contributions of this study. The estimates and the standard errors of the linear and quadratic trend before and after the respective dates for all provinces are displayed in \Cref{tab:trend-coef-china}. 

\begin{table}[!htb]
\centering
\caption{Estimates (and standard errors) for the trend function corresponding to best model for daily data for the Chinese provinces}
\label{tab:trend-coef-china}
\begin{tabular}{lccccc}
  \hline
  & & \multicolumn{2}{c}{before resp. date} & \multicolumn{2}{c}{after resp. date} \\
  Province & trend change & linear term & quad. term & linear term & quad. term \\
  \hline
  Anhui & 2020-02-11 & 1773.44 (1025.2) & -5.72 (46.9) & -1412.61 (824.7) & 1.67 (44.7) \\ 
  Beijing & 2020-02-11 & 0.47 (0.1)* & -0.02 (0)* & -0.96 (0.1)* & 0.02 (0)* \\ 
  Chongqing & 2020-02-09 & 0.99 (0.1)* & -0.05 (0)* & -1.45 (0.1)* & 0.05 (0)* \\ 
  Fujian & 2020-01-24 & -18.09 (3.8)* & 8.74 (1.8)* & 17.7 (3.8)* & -8.73 (1.8)* \\ 
  Gansu & 2020-03-05 & 0.23 (0.2) & -0.01 (0)* & 0.32 (0.2) & 0 (0) \\ 
  Guangdong & 2020-02-10 & 1.42 (0.4)* & -0.06 (0)* & -3.11 (0.4)* & 0.08 (0)* \\ 
  Guangxi & 2020-02-27 & 0.68 (0.1)* & -0.02 (0)* & -0.84 (0.1)* & 0.02 (0)* \\ 
  Guizhou & 2020-02-18 & 1.01 (0.1)* & -0.03 (0)* & -1.11 (0.1)* & 0.03 (0)* \\ 
  Hainan & 2020-02-24 & 0.63 (0.1)* & -0.03 (0)* & -0.98 (0)* & 0.03 (0)* \\ 
  Hebei & 2020-02-19 & 1.08 (0.2)* & -0.03 (0)* & -1 (0.2)* & 0.03 (0)* \\ 
  Heilongjiang & 2020-02-22 & 0.77 (0.1)* & -0.02 (0)* & -0.94 (0.1)* & 0.02 (0)* \\ 
  Henan & 2020-01-25 & -12.98 (1.7)* & 3.89 (0.5)* & 12.81 (1.7)* & -3.89 (0.5)* \\ 
  Hubei & 2020-01-30 & 0.69 (0.6) & -0.14 (0.1) & -0.53 (0.6) & 0.13 (0.1) \\ 
  Hunan & 2020-01-24 & -19.38 (2.6)* & 8.96 (1.2)* & 19.19 (2.6)* & -8.96 (1.2)* \\ 
  Inner Mongolia & 2020-02-20 & 0.85 (0.1)* & -0.04 (0)* & -1.21 (0.1)* & 0.04 (0)* \\ 
  Jiangsu & 2020-01-24 & -12.18 (2.1)* & 6.04 (0.9)* & 11.97 (2.1)* & -6.04 (0.9)* \\ 
  Jiangxi & 2020-02-19 & 0.55 (0)* & -0.02 (0)* & -0.55 (0)* & 0.02 (0)* \\ 
  Jilin & 2020-02-18 & 0.94 (0.1)* & -0.03 (0)* & -0.97 (0.1)* & 0.03 (0)* \\ 
  Liaoning & 2020-02-17 & 0.96 (0.2)* & -0.03 (0)* & -1.04 (0.2)* & 0.04 (0)* \\ 
  Ningxia & 2020-02-16 & 1.36 (0.4)* & -0.03 (0)* & -1.02 (0.3)* & 0.03 (0)* \\ 
  Qinghai & 2020-02-08 & 1.68 (0.2)* & -0.08 (0)* & -1.65 (0.1)* & 0.08 (0)* \\ 
  Shaanxi & 2020-02-23 & 0.67 (0.1)* & -0.03 (0)* & -1.02 (0.1)* & 0.03 (0)* \\ 
  Shandong & 2020-02-06 & 1.79 (0.4)* & -0.09 (0)* & -2.85 (0.4)* & 0.1 (0)* \\ 
  Shanghai & 2020-02-18 & 0.69 (0.3)* & -0.03 (0) & -1.13 (0.3)* & 0.03 (0)* \\ 
  Shanxi & 2020-01-26 & -12.26 (1.8)* & 2.92 (0.5)* & 11.91 (1.8)* & -2.91 (0.5)* \\ 
  Sichuan & 2020-02-29 & 0.63 (0.2)* & -0.02 (0)* & -0.96 (0.2)* & 0.02 (0)* \\ 
  Tianjin & 2020-02-23 & 0.72 (0.2)* & -0.02 (0)* & -0.84 (0.1)* & 0.02 (0)* \\ 
  Tibet & 2020-02-01 & 19.76 (0)* & -9.88 (0)* & -19.76 (0)* & 9.88 (0)* \\ 
  Xinjiang & 2020-02-19 & 0.7 (0.1)* & -0.02 (0)* & -0.85 (0)* & 0.02 (0)* \\ 
  Yunnan & 2020-02-21 & 0.8 (0.2)* & -0.02 (0)* & -0.84 (0.1)* & 0.02 (0)* \\ 
  Zhejiang & 2020-02-23 & 0.76 (0.3)* & -0.02 (0)* & -1.02 (0.2)* & 0.03 (0)* \\  
   \hline
\end{tabular}
\end{table}

\begin{figure}[!hbt]
\begin{center}
\includegraphics[width=\textwidth,keepaspectratio]{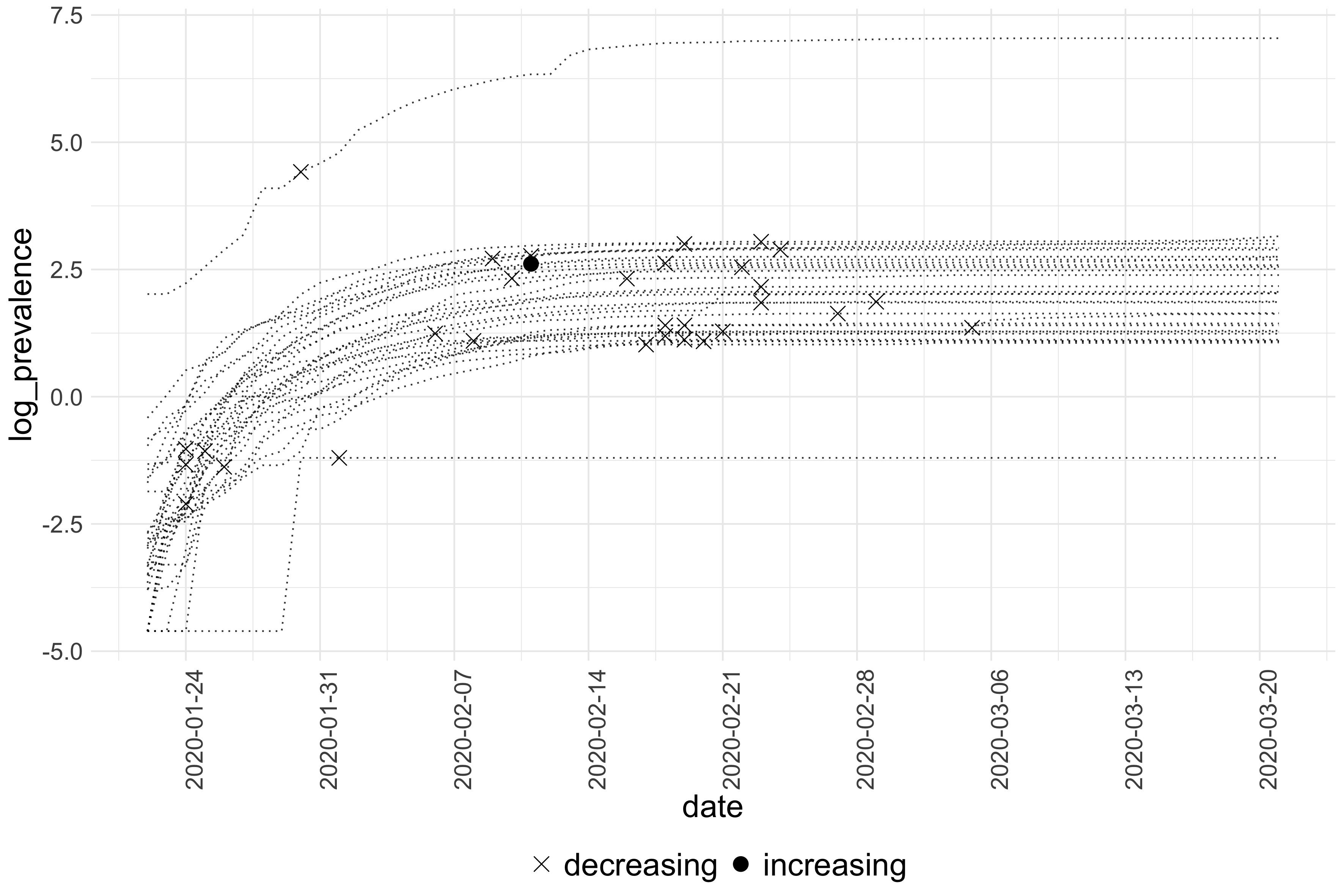}
\end{center}
\caption{log-prevalence of COVID-19, along with the direction of change in the trend function for log-incidence rate (plotted against the date when that change occurs), for different provinces in China. Anhui is the only province to show increasing behavior.}
\label{fig:trendchange-china}
\end{figure}

The corresponding dates vary between 24th January, 2020 to 5th March, 2020. But, most of them are concentrated in the week of 17th to 24th February. The obvious next question is whether the trend changes to halt the growth or to increase it. That aspect was explored using the estimated coefficients and fitted values for the trend function. In \Cref{fig:trendchange-china}, we show the log-prevalence of COVID-19 for different provinces. On each line in this graph, we have marked the date when the trend function for the log-incidence rate changes significantly. The direction of the change is also denoted in this graph, using $\times$ for decreasing and ${\bullet}$ for increasing. It is clear that all provinces but one (Anhui) observed a decreasing growth pattern after the respective date.

\subsection{Country-level analysis}
\label{subsec:global}

For this paper, we work with the COVID-19 cases from six countries - China, India, Iran, Italy, South Korea and the United States of America (US). These nations are considered for different reasons. China, being the epicenter of the outbreak, is definitely the most crucial country of the lot. While we have already performed a province-level analysis for China, we want to explore the efficiency of the modeling scheme on the country-level data as well. This would help us in confirming the conclusions from the previous section, thus establishing robustness of the model. 

Italy, Iran and South Korea were hit early with the outbreak. While they have continued to be three of the top ten countries with highest number of confirmed cases, they showed different patterns when it comes to fatality rates. Italy and Iran have 9.01\% and 7.55\% fatality rates respectively, both being much higher than the overall average. South Korea, on the contrary, has observed only 1.16\% fatality rate. Moreover, Korea have already adopted several screening and healthcare policies, and that may have helped them with the situation. It is also worth mention that Iran has a recovery rate which is on a higher side (37.05\%). This is second among all countries in consideration in our paper, bettered by only China who are reporting a 88.38\% recovery rate as of 21st March, 2020. 

Last but not the least, US and India got affected by the virus a bit later than the aforementioned countries and the growth rate is still believed to be at an early stage. Alarmingly, US is already at the third spot in the global list. India, meanwhile, is the only country in our study from outside the top ten. Despite being a nation of more than a billion people, it has reported only 330 confirmed cases, 5 having passed away till date. Refer to \Cref{tab:country-summary} for a detailed comparison of these numbers. 

\begin{table}[!htb]
\centering
\caption{Number of COVID-19 confirmed cases, corresponding number (percentage) of deaths and recovered cases, for a few countries and the World, as of 21st March, 2020.}
\label{tab:country-summary}
\begin{tabular}{lccc}
  \hline
  Country & Confirmed & Death & Recovered \\ 
  \hline
  China & 81305 & 3259 (4.01\%) & 71857 (88.38\%) \\ 
  Italy & 53578 & 4825 (9.01\%) & 6072 (11.33\%) \\ 
  US & 25489 & 307 (1.20\%) & 17 (0.07\%) \\ 
  Spain & 25374 & 1375 (5.42\%) & 2125 (8.37\%) \\ 
  Germany & 22213 & 84 (0.38\%) & 233 (1.05\%) \\ 
  Iran & 20610 & 1556 (7.55\%) & 7635 (37.05\%) \\ 
  France & 14431 & 562 (3.89\%) & 12 (0.08\%) \\ 
  South Korea & 8799 & 116 (1.16\%) & 1540 (17.50\%) \\ 
  Switzerland & 6575 & 75 (1.14\%) & 15 (0.23\%) \\ 
  United Kingdom & 5067 & 234 (4.62\%) & 67 (1.32\%) \\ 
  \hline
  India & 330 & 5 (1.52\%) & 23 (6.96\%) \\
  \hline
  World & 304524 & 12981 (4.26\%) & 91530 (30.06\%) \\
   \hline
\end{tabular}
\end{table}

Our main aim is to explore whether our model can provide valuable insight about the trend of the log-incidence of all of these countries. In addition, we also analyze the effect of lockdowns in Italy and China.

Looking at \Cref{tab:results-global}, we can say that the complete lockdown helped Italy reduce their number of affected cases significantly. The lockdown in China, which was considered to be 2nd February, 2020 (as major provinces implemented their restrictions by then), was not significant in the model. This is in line with what we saw in \Cref{subsec:china}. We also arrive at a similar conclusion about the nonstationarity and the most appropriate ARMA structure of the time series. South Korea showed a significant autocorrelation with prior three days. But, no other country has displayed the same beyond lag 1. It was further confirmed by the diagnostics we performed. The Box-Ljung test on the residuals turned out to be nonsignificant (p-values were greater than 0.05) for all countries. The prediction RMSE was good for most countries, except for Italy and US. 

\begin{table}[!htb]
\centering
\caption{Results from the best models corresponding to daily data of five countries}
\label{tab:results-global}
\begin{tabular}{lcccc}
  \hline
  Country & ARMA order & lockdown-effect & p-value (Box test) & RMSE (prediction) \\ \hline
  China & (0,1,0) & 0.45 (2.3) & 0.34 & 59.65 \\ 
  India & (0,1,0) &  & 0.15 & 31.35 \\ 
  Iran & (1,1,0) &  & 0.87 & 508.42 \\ 
  Italy & (1,1,0) & -1.82 (0.6)* & 0.70 & 1258.69 \\ 
  South Korea & (3,1,2) &  & 0.76 & 92.71 \\ 
  US & (0,1,0) &  & 0.77 & 3963.44 \\
   \hline
\end{tabular}
\end{table}

Next, we look at the dates when the trend function changed its curve significantly for these countries. The estimates and corresponding standard errors for the coefficients of the trend function of the models are presented in \Cref{tab:trend-coef-global}. 

\begin{table}[!htb]
\centering
\caption{Estimates (and std errors) for the trend function corresponding to best model for daily data for the five countries}
\label{tab:trend-coef-global}
\begin{tabular}{lccccc}
  \hline
  & & \multicolumn{2}{c}{before resp. date} & \multicolumn{2}{c}{after resp. date} \\
  Country & trend change & linear term & quad. term & linear term & quad. term \\ 
  \hline
  China & 2020-02-19 & 1.86(0.4)* & -0.05(0)* & -1.61(0.2)* & 0.04(0)* \\ 
  India & 2020-03-16 & -0.07(0)* & 0(0)* & -0.77(0.1)* & 0.02(0)* \\ 
  Iran & 2020-02-28 & 3.06(0.2)* & -0.22(0)* & -1.86(0.2)* & 0.2(0)* \\ 
  Italy & 2020-02-22 & -0.04(0.1) & 0(0) & 0.23(0.1)* & 0(0) \\ 
  South Korea & 2020-02-20 & 0.6(0.2)* & -0.02(0)* & 0.04(0.1) & 0.01(0)* \\ 
  US & 2020-03-10 & -0.07(0)* & 0(0)* & -0.03(0) & 0(0)* \\
   \hline
\end{tabular}
\end{table}

\begin{figure}[!hbt]
\begin{center}
\includegraphics[width=\textwidth]{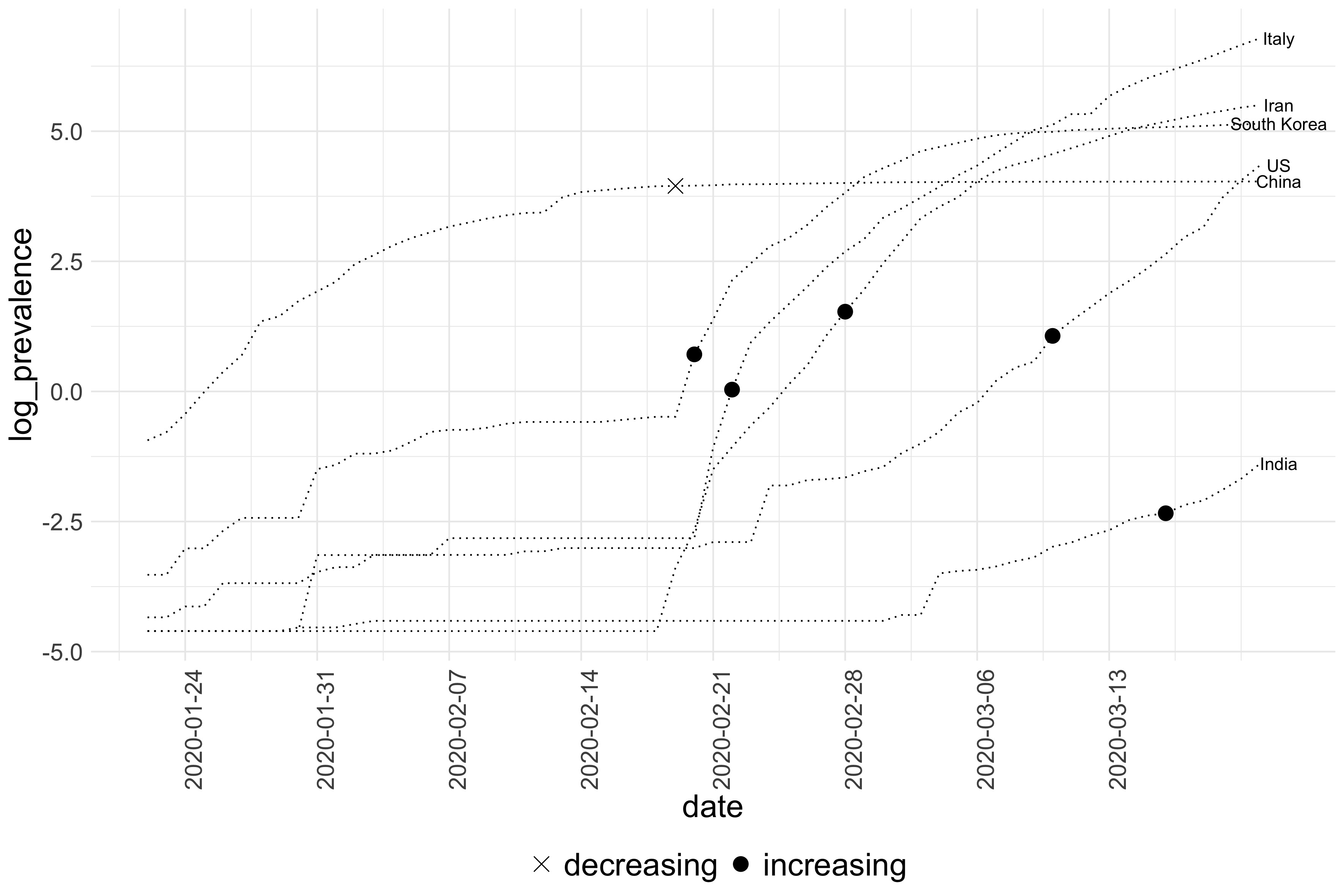}
\end{center}
\caption{log-prevalence of COVID-19, along with the direction of change in the trend function for log-incidence rate (plotted against the date when that change occurs), for the six countries in the study.}
\label{fig:trendchange-global}
\end{figure}

For China, it was 19th February after which the log-incidence started to decrease (refer to \Cref{fig:trendchange-global}). This is in accordance with what we observed in the previous section. 

Interestingly, contrary to what we noticed for China, the other countries showed a different pattern. For all five of these countries, the growth has not yet started to cease. Rather, it has been increasing more. Situations have been getting worse in Iran and Italy since the last week of February. US have seen the rise in confirmed cases after 10th March and it continues to be disastrous. India, albeit have managed to contain the pandemic so far, are seeing an increasing trend since 16th March and the number may rise drastically in coming days. For South Korea, on the other hand, the change in trend pattern came on 20th February. While the graph depicts that the situation has more or less stabilized over the last few days, statistically we cannot confirm it yet.

\subsection{Country-wise reproduction number}

\R was estimated for all of the above countries, using different serial interval distributions, as mentioned in \Cref{subsec:estimate-r0}. These results are reported in \Cref{tab:r0}. 

Earlier, \cite{chen2020mathematical} estimated \R to be 2.30 from reservoir to person and 3.58 from person to person. \cite{zhao2020preliminary} reported that the reproduction number varies between 2.24 to 3.58. Our estimates for all countries but the US fall in this range. For India, it is the lowest. But, that is also in accordance with \cite{world2020laboratory} who reported that \R for the human-to-human transmission ranges from 1.4 to 2.5. \R for US is surprisingly high and might be a reason to worry. As \cite{ridenhour2018unraveling} pointed out, \R may vary geographically ``owing to changes in environment, population structure, viral evolution, and immunity". Different healthcare and immigration policies might also be responsible in this regard. 

\begin{table}[!htb]
\centering
\caption{Estimated \R for different gamma distributions (mean $\pm$ standard deviation) of serial intervals}
\label{tab:r0}
\begin{tabular}{lccc}
  \hline
  Country & SARS ($8.4\pm 3.8$) & MERS ($7.6 \pm 3.4$) & average ($8.0 \pm 3.6$) \\ \hline
  China & 2.73 & 2.49 & 2.60 \\ 
  India & 1.43 & 1.40 & 1.42 \\ 
  Iran & 2.75 & 2.45 & 2.60 \\ 
  Italy & 3.14 & 2.87 & 3.00 \\ 
  South Korea & 2.25 & 2.04 & 2.14 \\ 
  US & 11.11 & 9.67 & 10.37  \\
  \hline
\end{tabular}
\end{table}

\section{Discussion}
\label{sec:discussion}

\subsection{Summary}

It is important to note that the time passed since the outbreak has been more than 2 months and we are yet to find a vaccine or an approved drug for treatment or pre-exposure prophylaxis for COVID-19. Thus, at the moment, epidemiological rather than pharmaceutical solutions need to be implemented, meanwhile promoting active research in finding a curative medicine. These measures need precise depiction of the ongoing disease and realistic growth predictions for proactive measures and continuous re-assessment of implemented regulations. This study endeavoured to sketch this very scenario using a statistical model taking in account the most recent data.

Using appropriate diagnostics, we confirmed that the model is able to capture the time-dependence pattern and the time trend of the disease-incidence. Combining all of the above analysis, it can be inferred that by the beginning of March, all provinces in China managed to contain the virus and consequently, the growth in the number of confirmed cases has been minimal afterwards. On the global scenario, all other countries are in a bad state. The growth rates of the incidence for these countries are increasing at the moment. As we analyzed the effect of lockdown, we found that for China, both on country-level and on province-level, it was not significant. Italy, however, was benefitted from going into complete lockdown.

We further want to mention that the spread of the COVID-19 across the globe has been described to occur in four stages. {\it Stage 1:} Cases are imported from affected countries. At this stage there is no spread of the disease locally. {\it Stage 2:} There is local transmission from infected persons, relatives or acquaintances of those who travelled abroad. During this time, source is known and it is easier to do contact tracing and self-quarantine. Countries like India are in stage 2. {\it Stage 3:} Community transmission has already begun. People with no prior history of travel or contact with affected persons test positive, meaning the source cannot usually be traced. We think US is one of the countries currently in Stage 3, as the number of cases has been growing substantially since 10th March, 2020. {\it Stage 4:} Phase of the infection where it reaches epidemic proportions, such that containing the spread is extremely difficult. According to our analysis, Italy, where the growth has an increasing trend since 22nd February and is yet to show a sign of decrease, is in stage 4. 

Clearly, the modeling approach gives an insight on what epidemiological stage a region is in. This has the potential to help in prompting implementation or change in approach to address COVID-19 in different countries.

\subsection{Limitations and Future work}

The main focus of this study has been on the trend pattern of the incidence of COVID-19. While we incorporated information on lockdown as best as we could, more detailed information on screening and testing measures, policies passed on different countries, effective preexposure drug profile access and robustness of healthcare services can be more effective in analyzing the data. Should that be available, one can easily include that information in the mean function of the model (cf. \cref{eqn:model}), given the flexible structure we consider.

On the other hand, in order to estimate \R in the most appropriate way, a detailed account of case-level data and case-to-case transmission data is needed. The same can be said about the estimation of serial interval as well. A future direction of this paper will be to use case-level data, once it is available, to estimate these two variables, which are crucial in an epidemiological study of an infectious disease.

\bibliography{deb_majumdar_covid}

\newpage

\section{Figures}
\label{sec:figures}

\end{document}